# Development of a Low-Level $^{37}$Ar Calibration Standard

R.M. Williams*, C.E. Aalseth, T.W. Bowyer, A.R. Day, E.S. Fuller, D.A. Haas, J.C. Hayes, E.W. Hoppe, P.H. Humble, M.E. Keillor, B.D. LaFerriere, E.K. Mace, J.I. McIntyre, H.S. Miley, A.W. Myers, J.L. Orrell, C.T. Overman, M.E. Panisko and A. Seifert

*Corresponding Author. Email: Richard.Williams@pnnl.gov

Pacific Northwest National Laboratory, PO Box 999, Richland WA 99352, USA

ABSTRACT

Argon-37 is an environmental signature of an underground nuclear explosion. Producing and quantifying low-level $^{37}$Ar standards is an important step in the development of sensitive field measurement instruments. This paper describes progress at Pacific Northwest National Laboratory in developing a process to generate and quantify low-level $^{37}$Ar standards, which can be used to calibrate sensitive field systems at activities consistent with soil background levels. This paper presents a discussion of the measurement analysis, along with assumptions and uncertainty estimates.

## 1. Introduction

An underground nuclear explosion is expected to produce significant quantities of $^{37}$Ar via neutron interaction with calcium in the surrounding soil (Aalseth et al., 2011; Riedmann and Purtschert, 2011). As a noble gas, $^{37}$Ar will diffuse through the soil without chemically binding and, given its 35-day half-life, will be present at detectable levels for months following the event. These qualities make $^{37}$Ar an attractive signature to be investigated during an on-site inspection (OSI) under the Comprehensive Nuclear-Test-Ban Treaty (CTBT). During such an OSI operation soil gas would be sampled and argon extracted for $^{37}$Ar analysis; the MARDS system is an example of field-deployable instrument for $^{37}$Ar analysis (Xiang et al., 2008). Ideally, such a system would be capable of measuring $^{37}$Ar in soil gas samples at or near the natural background, which is estimated to be in the range of 1-200 mBq.m$^{-3}$ of whole air (Riedmann and Purtschert, 2011). Pacific Northwest National Laboratory (PNNL) has developed a low-level $^{37}$Ar measurement capability based on an ultra-low background proportional counter (ULBPC) targeting natural background detection limits (Aalseth et al., 2013).

In order to fully optimize ULBPC detection of low $^{37}$Ar activities, a number of studies have taken place with the aim of characterizing the low-energy detection performance and operation of the ULBPC measurement system developed at PNNL (Seifert et al., 2013a; Seifert et al., 2013b). A significant effort has also been placed on developing a process for routinely producing low-level $^{37}$Ar samples and a method to accurately quantify the specific activity of those prepared samples so that they may be used to accurately calibrate the ULBPC detector efficiency.

PNNL has developed a low-level absolute gas counting capability based on the proven method known as length-compensated internal source proportional counting (Mace et al., 2013; Williams et al., 2013). The method has been adopted by many national metrology laboratories for determining the specific activity of a number of gaseous radionuclide standards (Unterweger, 2007). The capability developed at PNNL extends this useful absolute gas counting technique to quantify materials at much lower activities, such as $^{37}$Ar, for use as low-level standards.

A brief description of the $^{37}$Ar production method is given below along with a discussion of two distinct sets of measurements performed on different aliquots of the same starting $^{37}$Ar material separated in time by several weeks. An estimate of the uncertainty in the estimated specific activity is provided based on the ISO Guide to the Expression of Uncertainty in Measurements (GUM) (ISO, 1995).

## 2. Argon-37 Production

For background purposes, a brief description of the preparation of $^{37}$Ar is given here. Leveraging the same $^{40}$Ca(n,α)$^{37}$Ar reaction which produces $^{37}$Ar in the soil following a nuclear explosion, a small sample of calcium carbonate was irradiated with 14 MeV neutrons from a commercial D-T generator. For a typical production run, a small borosilicate glass vial containing 1 gram of calcium carbonate is irradiated for up to 4 hours; then short-lived activation products are allowed to decay during a 48 hour post-irradiation period before the sample is processed further. The 14-MeV neutron flux near the vial is estimated to be on the order of $1\times10^{9}$ n.cm$^{-2}$.s$^{-1}$.

Prior to irradiation, the calcium carbonate is sealed in the 20-mL glass vial using a standard cap fitted with a polymer septum gasket. Helium gas is repeatedly flushed through the headspace of the vial to remove most of the residual ambient air as well as to act as a carrier gas to facilitate transfer of the trace amounts of $^{37}$Ar produced during the irradiation. Following the irradiation (and cool down) a glass syringe fitted with an integrated valve is used to extract several cubic centimeters of the $^{37}$Ar-spiked helium gas from the headspace. The syringe is then attached to a gas manifold and the helium sample is expanded into a section where approximately 20 mL of ultra-high-purity stable argon is added. The argon/helium sample is then exposed to a heated getter that removes trace residual air constituents. The purified sample is then exposed to a u-shaped trap containing a few milligrams of activated charcoal submerged in liquid nitrogen, in which argon is preferentially adsorbed over helium. After several minutes the trap is isolated from the getter section and warmed. The argon sample (with percent levels of helium) is volumetrically expanded into a 50-mL container, which is backfilled to 5 bar with P10 counting gas and allowed to equilibrate for 24 hours before further processing. At this stage the composition of the sample gas is nearly the same as P10; slightly more argon is present from the initial dilution with stable argon but this does not affect the measurement. This spiked P10 sample represents the 'starting' material from which aliquots are drawn for either quantification or for detector calibration. Dilution factors prior to this step are not used in calculations of specific activity and the mole fraction of argon present is assumed to be 90% (the processing method has recently been modified so that it is more quantitative up to this point).

The 250 mL of spiked P10 is volumetrically expanded into a manifold fitted with 14 separate sample containers, whose individual physical volumes had previously been measured (each container very close to 6 mL); because of the volume ratios the final pressure in the entire system was slightly below 1 bar. As the valves for each of the 14 containers are closed the pressure and temperature are noted and used to calculate the P10 gas volume (at STP). Two of these containers were selected for specific activity quantification (the first and last container out of 14 containers produced were selected in order to observe any potential difference due to sample preparation; none were observed). One goal of quantifying these two samples was to test the assumption that the sample preparation method was producing samples with the consistent specific activity. As the method for producing $^{37}Ar$ was still being optimized, the sample was first transferred into the largest of three unequal-length ULBPC detectors (100 mL) in order to conduct an initial assessment of the activity. The detector and the 6-mL sample were attached to a loading manifold. From there the sample was quantitatively transferred into the 100 mL detector followed by P10 count gas, bringing the final pressure of the detector to 7 bar. The detector was then taken to PNNL's shallow underground laboratory (Aalseth et al., 2012) for initial assessment of the sample to ensure the $^{37}Ar$ production and sample transfer were successful. After several days of counting, the sample was ready for transfer into the other detectors for absolute determination of the specific activity.

## 2.1 Length-Compensated Measurements

Length-compensated internal-source proportional counting has been used and refined since the early demonstrations by Anderson and Libby (Anderson et al., 1947). The technique combines data from multiple counters that are identical in every way except for their length, to remove the so-called end effect. The end effect is a phenomenon that reduces the overall detection efficiency and occurs near the ends of detectors, where the electric fields are non-uniform and difficult to accurately model. Investigating the differential count rate between detectors of unequal length neutralizes the end effect and the overall efficiency then becomes a function of phenomena such as threshold and wall effects; both of which can be more accurately estimated through modelling, experiments or a combination of both. The length-compensated method has been used as a primary measurement method in the determination of gaseous radionuclides by a variety of national metrology institutes (Makepeace et al., 1998; Picolo et al., 1998). A variant of this method has been developed at PNNL for quantification of low-level gaseous radionuclides by employing materials and methods used in the production of ULBPCs (Mace et al., 2013; Williams et al., 2013). The PNNL absolute gas counting system consists of a gas handling and mixing manifold (used to quantitatively blend samples with P10 counting gas and transfer this mixture to the detectors); a set of three unequal-length ULBPCs, which are filled above ground then taken to PNNL's shallow underground laboratory for counting; and a set of four counters fabricated from oxygen free high conductivity (OFHC) copper. The set of four OFHC counters is permanently connected to the gas blending system and also housed in an active/passive shield for counting in an above-ground laboratory. For the work presented here, the ULBPC detectors were filled above ground and taken to the shallow laboratory for measurement.

After the initial assessment has shown that the sample has measurable activity and the $^{37}$Ar signatures are consistent with previous measurements (Aalseth et al., 2013) the detector is returned from the shallow underground counting laboratory and attached a transfer manifold. Also attached to the same manifold are the other two shorter detectors that make up ULBPC absolute counting set. The 7-bar sample in the 100-mL detector is expanded into the other two detectors while the temperature and pressures are monitored (the final pressure is 2.5 bar); no additional P10 is added. The volumes of these three ULBPC detectors have been measured as 67.56 (±0.14) mL, 84.37 (±0.22) mL and 100.59 (±0.21) mL, respectively (numbers in parenthesis refer to the expanded uncertainty (k=2) of the measurement). All three detectors are placed on a copper baseplate housed in an insulated enclosure; the baseplate facilitates thermal equilibrium between the three detectors. Calibrated thermistor temperature probes measure the temperature (Fluke Model 1529) of the detectors at three locations and all show excellent uniformity. The pressure is measured using a quartz-resonator-based commercial transducer with excellent precision and accuracy (Paroscientific Model 6000). Once the pressure and temperature have equilibrated, the detectors are removed from the manifold and taken back to the shallow underground laboratory for extended measurements. Prior to these measurements, the background count rates for these three detectors in the shallow counting laboratory were determined over a range of pressures (Mace et al., 2013) and found to be <10 counts per day in the $^{37}$Ar region of interest (ROI).

## 3. Results

The first sample was measured for 12 days; the beginning of the measurement acquisition was 18 days following sample extraction from the irradiated sample via and initial processing. The measurements related to the second sample were started 60 days after initial sample preparations; due to the lower activity of this second sample it was counted for 39 days. The count rate was corrected for the actual sample decay during the counting interval, since it was comparable to the 35-day half-life of $^{37}$Ar.

The data from each detector was used to derive the net count rate following the analysis given by Aalseth et al. (2013); this involves fitting the observed 2.8-keV combined Auger and x-ray spectral feature. Figure 1 is a compilation of the results for the first set of measurements, showing the energy histograms and the peak fit. The data for the second set of measurements is very similar except for reduced count rate due to the decay of the sample.

To estimate the specific activity from a dataset from a set of counters with unequal lengths, it is customary to compute the ratio of the differential net count rate to the differential gas volume, and then correct for wall effects (those events that produce β emission that strikes the counter wall before producing sufficient electrons for detection) and threshold effects (decay events producing pulses that fall below the detection threshold). These effects can be significant for certain radioisotopes, such as $^{39}$Ar, which possesses a broad β distribution with a relatively high end-point energy (565 keV); however, for $^{37}$Ar the emission is limited to narrow, low-energy range around 2.8 keV. Simulations of electron transport at these pressures suggest that a very high percentage (>99.9%) of electrons are stopped by the count gas before striking the walls. Additionally, due to the peaked distribution of observed 2.8-keV events, it is assumed that 100% of these events occur above the low-energy threshold (set at 1 keV). As a result, the ratio of the differential count rate to the differential gas volume is very close to the actual specific activity of the sample, with the exception of any correction for branching ratio. The 2.8-keV $^{37}$Ar feature observed in this work is due to a combination of the 81.5% of decays that yield Auger electrons summing to 2.82 keV and the 8.7% of decays that yield x-rays of the same energy for a combined branching ratio of 90.2% (Aalseth et al., 2013; Barsanov et al., 2007).

A convenient way to represent and analyze the data from the set of three unequal-length counters is to plot the argon gas volume (mL at STP) versus the decay rate (Bq) and fit the data to a linear model. Figure 2 is the data from both sets of measurements. The decay rate is simply the net count rate divided by the expected branching ratio (0.902). The argon gas volume is calculated using

$$Vol_{STP} = \frac{P(kPa)}{P_0} \frac{T_0}{T(K)} Vol_{detector}(mL) \cdot mf \qquad (1)$$

with the IUPAC STP definitions of $P_0$ (100 kPa) and $T_0$ (273.15 K). The physical volumes of the ULBPC detectors were measured at PNNL using a dedicated apparatus with a known reference volume. The total fill pressures and temperatures for measurements 1 & 2 were 239.834 kPa at 295.14 K and 241.34 kPa at 294.470 K, respectively. The mole fraction of argon in the count gas is assumed to be 0.90. The data from both sets of measurements are shown in Figure 2, sources and magnitudes of uncertainty will be discussed in the next section. The difference in slope between the two sets of data is the result of sample decay that occurred between the first and second sets of measurements. Although there are only three data points per measurement set, the linearity is very good, suggesting the reproducibility in detector construction is high; a requirement for elimination of the end effect. The dashed lines above and below each set of data represent the estimated slope and intercept with their estimated uncertainty added or subtracted to provide a visual reference of the uncertainty. An advantage of plotting the data in this manner is that the vertical-axis intercept provides a direct estimate of the actual volume of gas that produces no net counts in the detector (under those operating conditions). This is a very important result as it provides a direct measure of the active versus non-active volume of the detector, which can help in assessing the overall efficiency and performance of the design. In this case, the intercept represents a gas volume; in order to convert this to a physical volume, equation 1 must be rearranged to solve for detector volume:

$$Inactive\ Vol_{detector}(mL) = \frac{P_0}{P(kPa)} \frac{T(K)}{T_0} \frac{1}{mf} \cdot GasVol_{STP}(mL)\ . \qquad (2)$$

For this work with $^{37}$Ar and operating near 2.7-bar pressure, the 'dead volume' of the detector is between 22 and 25 mL and is consistent with the design and the low energy of the detected radiation. It is expected that the active or fiducial volume of the detector will be a function of the operating conditions and the isotope being investigated. Table 1 summarizes the results from the linear regression of the data shown in Figure 2 and also provides an estimate of the decay-corrected $^{37}$Ar specific activity on the date the original dilutions occurred. The production, extraction, and subsequent dilution of the $^{37}$Ar from the original irradiated vial yielded an air equivalent $^{37}$Ar specific activity of >1100 mBq·m$^{-3}$ of air. This level is up to 1000 times greater than the lowest expected soil background; however, given the 35-day half-life of $^{37}$Ar, it is a useful starting value, especially if fresh production runs occur semi-annually. For example, if several productions were to occur over 12 months, a range of specific activities could be available that span a wide range of expected soil background levels, providing a useful standard for calibration and/or validation of $^{37}$Ar measurement systems.

## 4. Estimated Uncertainty Discussion

The uncertainty estimate for these measurements involves developing a comprehensive data analysis model that includes uncertainty. The framework for uncertainty model was based upon approaches and suggestions outlined in the ISO GUM (ISO, 1995). To facilitate the integration of data analysis and uncertainty estimation, a model was developed within a software tool known as the GUM Workbench (Metrodata-GmbH, 2012), which propagates uncertainty along with basic calculations.

The uncertainty model consists of three significant elements: 1) uncertainty associated with the gas volume for each of the three detectors, 2) uncertainty associated with calculating the decay rate for each detector, and 3) estimating the uncertainty in the slope and intercept of the linear regression analysis. Integrating these three steps in the overall uncertainty model allows for correlated uncertainties to be accounted for, thus providing a best estimate of the overall uncertainty. Equation (1) is used in the analysis of the argon gas volume for each of the three detectors; the uncertainties in the temperature, pressure, mole fraction and detector volumes are included. The basic equation used in the estimate of the decay rate is given by

$$DecayRate(Bq) = \frac{\lambda_{37Ar}}{(1-exp(\lambda_{37Ar}\, t_{live}))} \frac{Net\ Counts}{\varepsilon_{wall-threshold} \cdot BR}, \qquad (3)$$

where the uncertainty in the branching ration (BR), wall and threshold effects, decay constant, live time and net counts are included. A detailed discussion of the $^{37}$Ar peak fitting and uncertainty estimate of the net count rate is given in Aalseth et al. (2013). Equations 1 and 3 are used to produce the three data points and associated uncertainty bars shown in Figure 2. Standard linear regression formulas are used to compute the slope and intercept from these three data points:

$$m = \frac{N\sum_{i=1}^{N} x_i y_i - \sum_{i=1}^{N} x_i \sum_{i=1}^{N} y_i}{N\sum_{i=1}^{N} x_i^2 - \left[\sum_{i=1}^{N} x_i\right]^2}\ ;\ b = \frac{\sum_{i=1}^{N} y_i - m\sum_{i=1}^{N} x_i}{N}, \qquad (4)$$

where $x_i$ are the calculated decay rates, $y_i$ are the calculated gas volumes and N = 3. Using this approach, the uncertainties in the slope and intercept can be determined easily using standard uncertainty propagation; in addition, correlations between uncertainties within each ($x_i$, $y_i$) pair are also easily accounted for. For example, the uncertainty in the measure of the physical volume of each detector is considered correlated because they were measured on the same apparatus using a standardized protocol. The uncertainty for the volume measurements was found to be dominated by the Type B uncertainty (related to systematic uncertainty or bias) of the pressure transducer.

An uncertainty budget for first measurement cycle is tabulated in Table 2. The overwhelming source of uncertainty, not surprisingly, originates in the counting statistics, which are considered type A using the GUM method of categorizing uncertainty, as they are statistical in nature. If a higher-activity sample were measured or if the counting interval were increased, the overall uncertainty would be reduced. It is also worth noting that this is the uncertainty in the specific activity referenced to the beginning of the counting interval; if a large decay correction is applied to this value and the correction spans multiple half-lives, the uncertainty in the half-life may become significant. For example, a 1% relative uncertainty in half-life becomes a 5% relative uncertainty in decay correction if the interval is 5 half-lives; therefore it is important to use the most accurate value available.

## 5. Conclusion

This paper has presented a method for producing modest amounts of $^{37}$Ar from neutron-irradiated $CaCO_3$. The process described yields multiple samples of usable activity from a single irradiation for the purpose of calibrating internal-source proportional counters. The material was analyzed using PNNL's low-level absolute gas counting capability based on a set of three unequal-length ultralow-background proportional counters. In order to test the consistency of the sample preparation method two different samples of the same starting material were measured using the length compensated approach; after decay correcting to a common reference date, both sets of measurements agree within the stated uncertainty. The $^{37}$Ar specific activity at the time of initial sample preparation is roughly 1100 mBq/m$^3$ air, which is above the expected soil background levels associated with possible OSIs under the CTBT; further, given the isotope's 35-day half-life, soil background levels would be available within a matter of months after preparation.

Acknowledgements

This document is PNNL-SA-109332. The research described in this paper is part of the Ultra-Sensitive Nuclear Measurements Initiative at Pacific Northwest National Laboratory. It was conducted under the Laboratory Directed Research and Development Program at PNNL, a multiprogram national laboratory operated by Battelle for the U.S. Department of Energy. Pacific Northwest National

Laboratory is operated for the U.S. Department of Energy by Battelle under contract DE-AC05-76RL01830.

Figure Captions

Figure 1

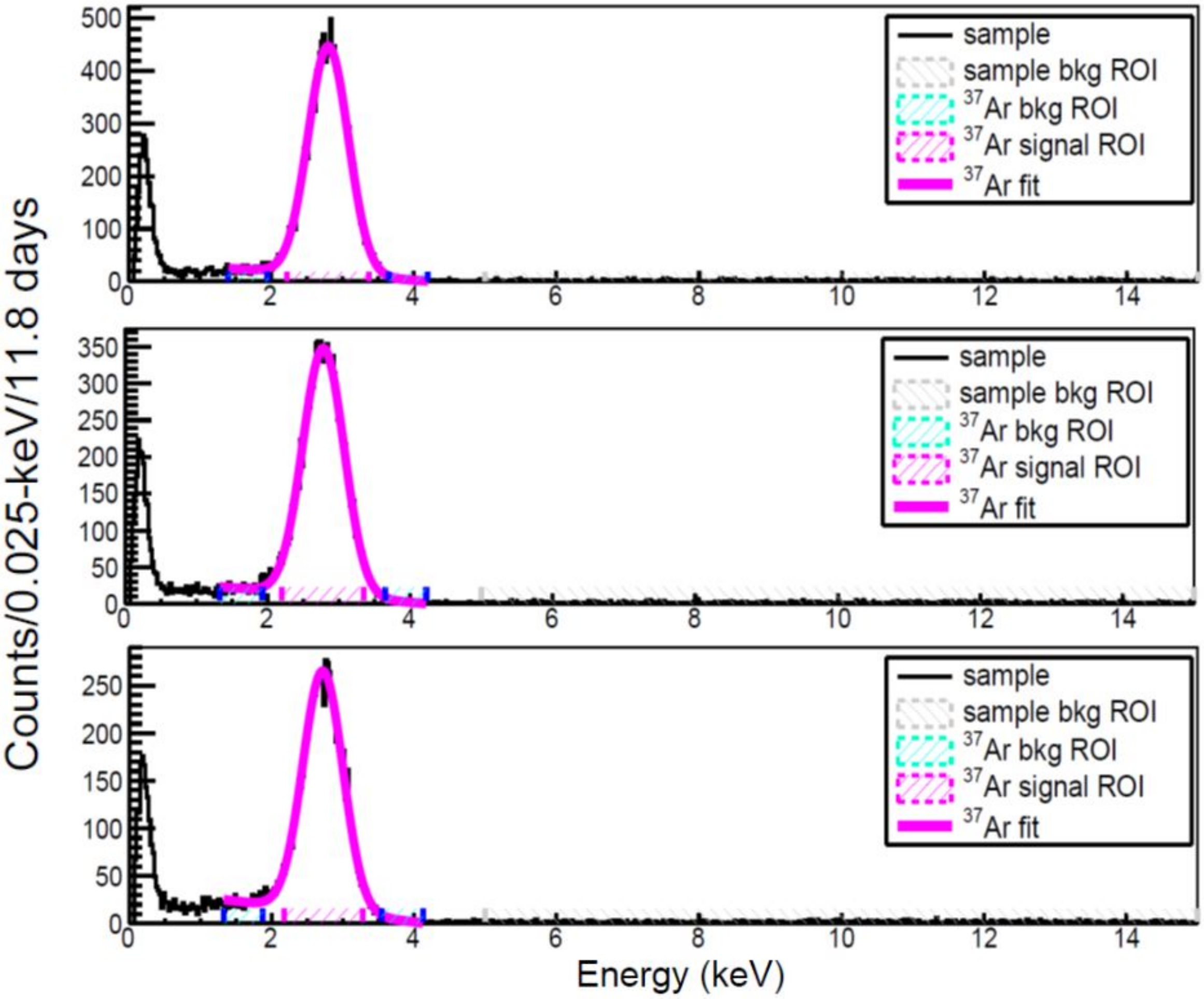

Figure 1: Compilation of energy histograms from the three ULBPC measurements of $^{37}$Ar. These data were from the first of two measurements made on the same starting material. The solid magenta lines represent a Gaussian fit of the 2.8 keV feature used in the estimate of gross and net counts.

Figure 2

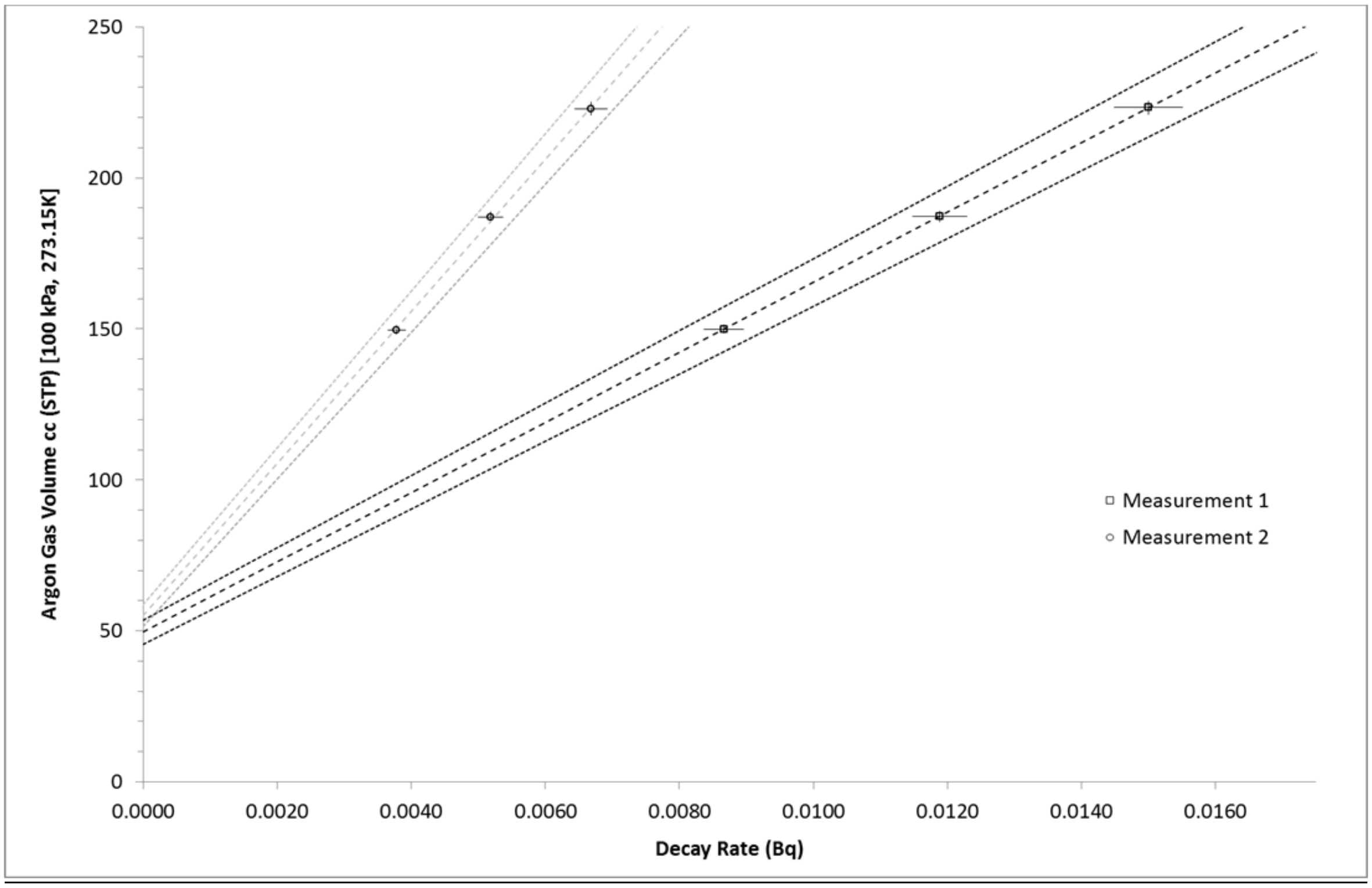

Figure 2: Data and fits from both sets of unequal length ULBPC $^{37}Ar$ measurements. Uncertainty bars and dashed lines represent the k=2 expanded uncertainty in the individual data points and linear regression, respectively.

Table Captions

Table 1:

| | Days Counted | Slope ($mL.Bq^{-1}$) | Intercept (mL) | Specific Activity ($Bq.mL^{-1}$) at beginning of acquisition | Specific Activity ($Bq.mL^{-1}$) at 22-04-2014 19:00 UTC | Air Equivalent* $^{37}Ar$ Specific Activity ($mBq.m^{-3}$) at 22-04-2014 19:00 UTC |
|---|---|---|---|---|---|---|
| Measurement 1 | 11.8 | 11579.43 (712/6.1%) | 49.62 (8.0/16.1%) | $86.36\times10^{-6}$ ($5.3\times10^{-6}$/6.1%) | $121.27\times10^{-6}$ ($7.5\times10^{-6}$/6.1%) | 1132.7 |
| Measurement 2 | 38.7 | 25214.32 (1460/5.8%) | 54.98 (7.2/13.1%) | $39.66\times10^{-6}$ ($2.3\times10^{-6}$/5.8%) | $127.44\times10^{-6}$ ($7.4\times10^{-6}$/5.8%) | 1190.3 |

* Assumes 0.934% argon in ambient air (by volume)

Table 1: Summary of the linear regression results for the data shown in Figure 2. Values in parenthesis represent expanded uncertainty (k=2) for each value.

Table 2:

| Quantity | Value | Standard Uncertainty | Sensitivity Coefficient | Uncertainty Contribution ($Bq.mL^{-1}$) | Index (%) |
|---|---|---|---|---|---|
| Net Counts (det1) | 7101 | 96.3 | $-1.70\times10^{-9}$ | $-1.60\times10^{-6}$ | 36.9 |
| Net Counts (det2) | 9744 | 110 | NA | $2.40\times10^{-9}$ | 0.0 |
| Net Counts (det3) | 12290 | 121 | $1.70\times10^{-9}$ | $2.00\times10^{-6}$ | 56.8 |
| Half-life (seconds) | 3027456 | 864 | $-3.20\times10^{-12}$ | $-2.80\times10^{-9}$ | 0.0 |
| Live-time-1 | 1018662.7 | 10 | $1.00\times10^{-10}$ | $1.00\times10^{-9}$ | 0.0 |
| Live-time-2 | 1018663.5 | 10 | $-1.80\times10^{-12}$ | $-1.80\times10^{-11}$ | 0.0 |
| Live-time-3 | 10186635 | 10 | $-1.80\times10^{-10}$ | $-1.80\times10^{-9}$ | 0.0 |
| Branching Ratio | 0.902 | 0.002 | $-9.60\times10^{-5}$ | $-2.30\times10^{-7}$ | 0.8 |
| Wall & Threshold Effect | 1 | 0.005 | $-8.60\times10^{-5}$ | $-4.30\times10^{-9}$ | 2.6 |
| Detector 1 Volume (mL) | 67.560 | 0.074 | $2.60\times10^{-6}$ | $2.00\times10^{-7}$ | -0.3 |
| Detector 2 Volume (mL) | 84.370 | 0.093 | $-3.30\times10^{-8}$ | $-3.00\times10^{-9}$ | 0.0 |
| Detector 3 Volume (mL) | 100.590 | 0.111 | $-2.60\times10^{-6}$ | $-2.90\times10^{-7}$ | 0.4 |
| Pressure (kPa) | 239.834 | 0.239 | $-3.60\times10^{-10}$ | $-8.60\times10^{-8}$ | 0.1 |
| Temperature (K) | 295.140 | 0.05 | $2.90\times10^{-7}$ | $1.50\times10^{-8}$ | 0.1 |
| Mole Fraction | 0.900 | 0.005 | $-9.60\times10^{-5}$ | $-4.30\times10^{-7}$ | 2.6 |
| **37Ar Specific Activity (05-09-2014 16:00 PST)** | **$86.36\times10^{-6}$ ($Bq.mL^{-1}$)** | | | | |
| **Standard Uncertainty** | **$2.65\times10^{-6}$ ($Bq.mL^{-1}$)** | | | | |
| **Expanded Relative Uncertainty (k=2)** | **6.1%** | | | | |

Table 2: Quantities and associated uncertainties used to estimate the total expanded uncertainty of $^{37}Ar$ specific activity for the first of two absolute gas counting measurements.